\documentclass[%
 reprint,
 amsmath,amssymb,
 aps,
]{revtex4-2}

\usepackage{graphicx}
\usepackage{dcolumn}
\usepackage{bm}
\usepackage{float}
\usepackage{subfigure}
\usepackage{color,soul}

\begin{document}

\preprint{APS/123-QED}

\title{Machine learning potential as a guide for eutectic in ultra-refractory multicomponent ceramics}

\author{V.~E.~Valiulin}
\email{valiulin@phystech.edu}
\affiliation{Moscow Institute of Physics and Technology (National Research University), Dolgoprudny 141701, Russia}
\affiliation{Institute for High Pressure Physics, Russian Academy of Sciences, Moscow (Troitsk) 108840, Russia}

\author{A.~V.~Mikheyenkov}
\affiliation{Moscow Institute of Physics and Technology (National Research University), Dolgoprudny 141701, Russia}
\affiliation{Institute for High Pressure Physics, Russian Academy of Sciences, Moscow (Troitsk) 108840, Russia}

\author{N.~M.~Chtchelkatchev}
\affiliation{Moscow Institute of Physics and Technology (National Research University), Dolgoprudny 141701, Russia}
\affiliation{Joint Institute for Nuclear Research, Dubna 141980, Russia}

\author{E.~A.~Levashov}
\affiliation{National University of Science and Technology ''MISIS``, Moscow 119049, Russia.}


%

\date{\today}

\begin{abstract}
\textbf{The experimental determination of eutectic points is a long-established and widely used technique, but it is generally only practical for systems with relatively low melting points. Many modern, promising materials, however, are ultra-refractory, with melting points exceeding 3000 K. For these systems, conventional melting experiments become prohibitively expensive and technically challenging. Advanced AI modeling can serve as a powerful precursor to guide successful experimentation in such cases. This work proposes a novel criterion for determining the eutectic point concentration in ultra-refractory alloys. The approach is verified using the Ti-B-C system—the most thoroughly studied three-component refractory system to date. The core of the algorithm is a machine-learning interatomic potential, based on a neural network, which achieves accuracy comparable to ab initio methods. Crucially, the algorithm operates effectively in the liquid phase, eliminating the need for information about the solid alloy's crystalline structure to estimate eutectic points.}
\end{abstract}

\maketitle


\section{\label{sec:level1}Introduction}

The creation of composites with exceptional thermomechanical properties, phase stability, mechanical strength and resistance to thermal shocks and aggressive environments has determined the scientific direction --- ultra-refractory materials based on borides and carbides of metals of groups IV-VI of the Periodic Table of Elements \cite{Chen16_JACS}. Such materials have high values of melting point, hardness, thermal conductivity, electrical conductivity, resistance to oxidation, wear, corrosion \cite{Eswara23_NRM,Schrei22_MB,Fahren17_SM}. Promising is the creation of solid substitution solutions that combine two or more metals in the metallic sublattice of a boride and/or carbide \cite{Otani09_JAC,Kurbat19_CI,Voroti22_JAC,Neuman21_JECS,Guo18_CI,Ivashc21_MCP,Hassan19_IJRMHM}, as well as the formation of eutectic and pseudo-eutectic ceramics \cite{Chen16_JACS,Remark1,Remark2}.

Various methods are used to obtain refractory compounds. In particular, self-propagating high-temperature synthesis (SHS) is an effective one due to its high productivity, low energy consumption, scalability \cite{Levash17_IMR}, and the properties are determined by the macrokinetic parameters of combustion and the mechanism of structure formation of the synthesis products \cite{Borovi17_Book,Morsi12_JMS,Aruna08_COSSMS}. Unsurpassed mechanical properties and oxidation resistance \cite{Voroti19_CMT} have been achieved in comparison with similar compositions obtained by sintering mixtures of individual carbides or borides \cite{Voroti19_JAC,Kurbat21_CI,Kurbat18_CI}. The authors of Refs. \cite{Zaitse23_IJSPHTS,Zaitse24_IJSPHTS} obtained ultrafine-grained ceramics based on borides, carbides (Ref.\cite{Iatsyu18_JECS}) and the HfB$_2$-HfC eutectic.

Structural engineering offers opportunities for optimizing the properties of ultra-refractory materials \cite{Ordan87_SPMMC,Zhao14_CI,Rezaie07_JMS,Zhang11_JECS,Sugiya08_MT,Nino11_MT,Zhou2025,Liu2025}. Eutectic composites are formed by invariant liquid–solid phase transformations and form composite microstructures with lamellar or rod-shaped morphology, while thermodynamic compatibility of the phases that make up the eutectic improves their characteristics \cite{AshbroO77_JACS,LLorca06_PMS}. Notably, eutectic ceramics exhibit increased mechanical, thermal, and chemical stability \cite{Brewer04_AM,Brewer03_JACS}.

It should be noted that, at grain sizes below the critical value, eutectic ceramics, consisting of brittle, highly hard, oxygen-free compounds, exhibit microplasticity that is not characteristic of the individual compounds. During crystallization, eutectics form interphase boundaries that, due to their unique structural configurations, have minimal energy. In MeC–MeB$_2$ systems, consisting of a mixture of phases with cubic and hexagonal crystal structures, a virtually ideal (or coherent) phase boundary arises due to the superposition of the \{111\}MeC and {001}MeB$_2$ planes. Thus, the increased plasticity can be due to the movement of dislocations from the bulk of the grains to the interface \cite{Padern92_SPMMC}.

Technical requirements for application objects often require alloying simple eutectics by introducing a 4th or even 5th element. In this case, eutectic compositions can be obtained in systems between solid solutions of the type (Me$^1_x$,Me$^2_y$)B$_2$ -- (Me$^1_x$,Me$^2_y$)C, and a change in the stoichiometric coefficients (x, y) leads to a shift in the eutectic points. Clarification of the composition of the carbide-boride eutectic is an independent labor-intensive task. One of the known methods for calculating eutectic points in three- and four-component metal systems is CALPHAD \cite{Zhang22_JPED}.

The experimental investigations of new ultra-refractory compositions are an extremely expensive and time-consuming process~ \cite{Sopata19_JAC, Silves23_JAC, Shaikh23_JAC}. Therefore, numerical algorithms are valuable to narrow down the search area for the eutectic composition. Additionally, the algorithm must operate in the liquid phase of the melt, as the details of the pre-eutectic and post-eutectic crystal structures of the new alloy are initially unknown.

\begin{figure*}[tph]
  \centering
  \mbox{
    \subfigure[] {\includegraphics[height=0.23\textheight]{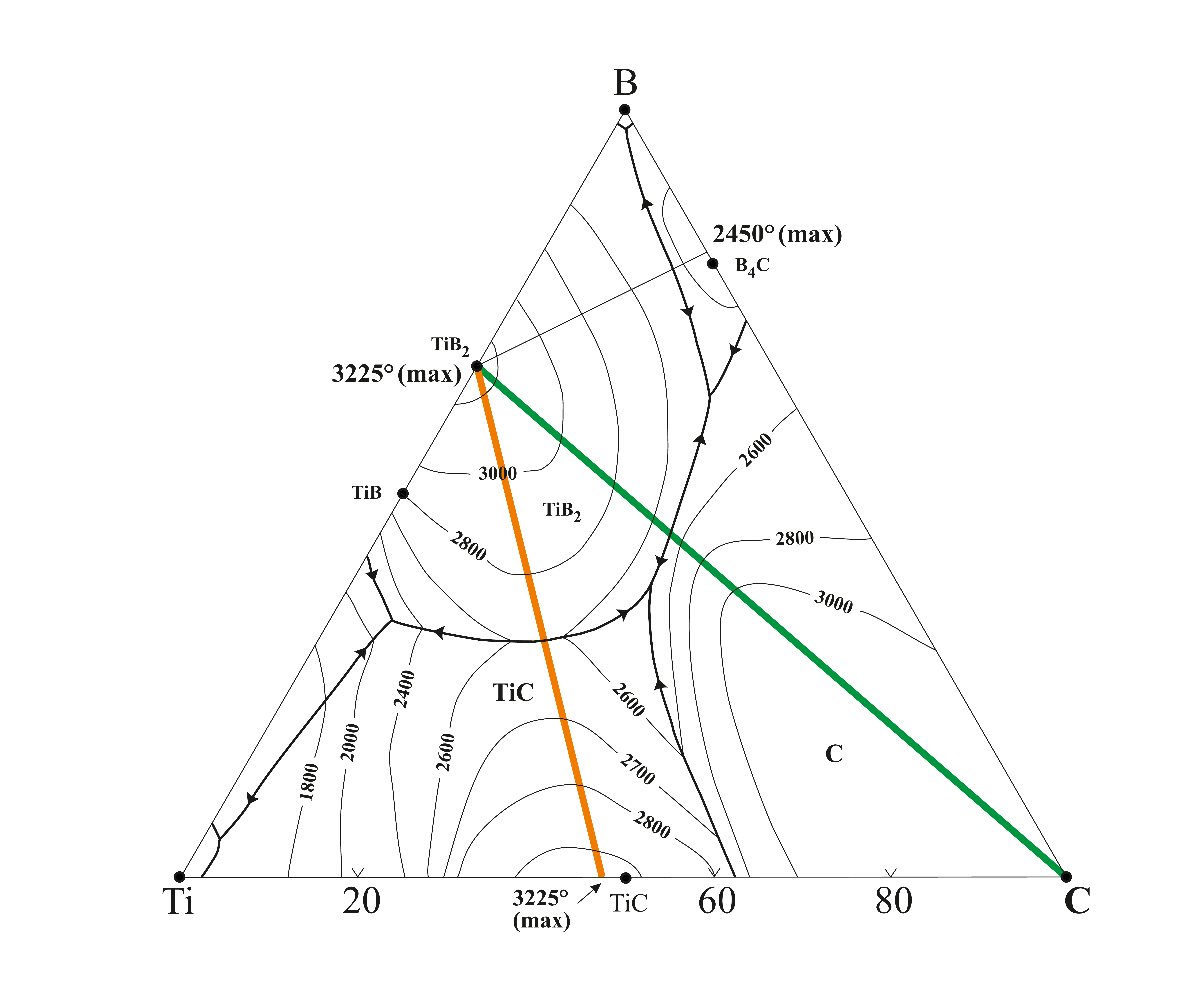}}
    \subfigure[] {\includegraphics[width=0.3\textwidth]{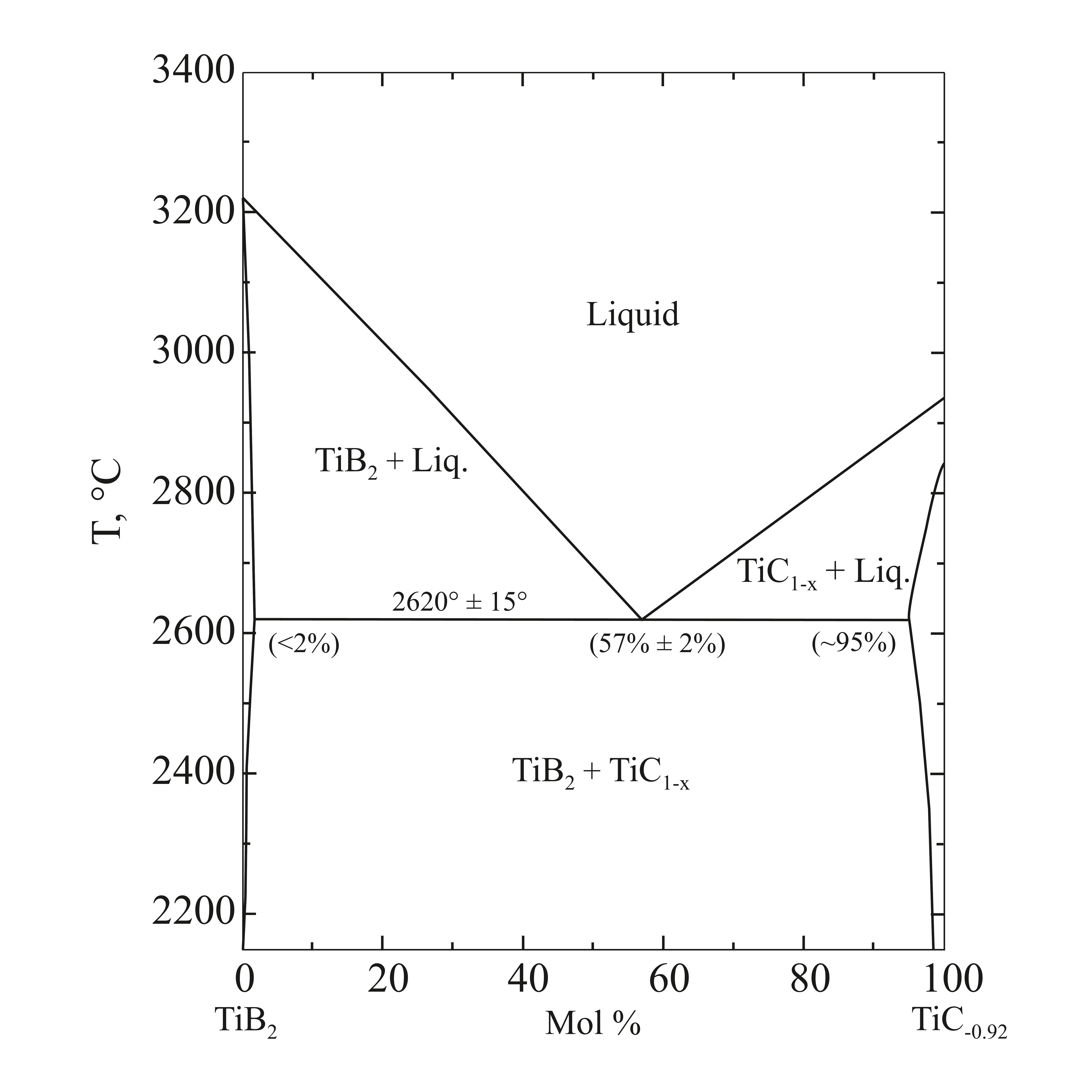}}
    \subfigure[] {\includegraphics[width=0.3\textwidth]{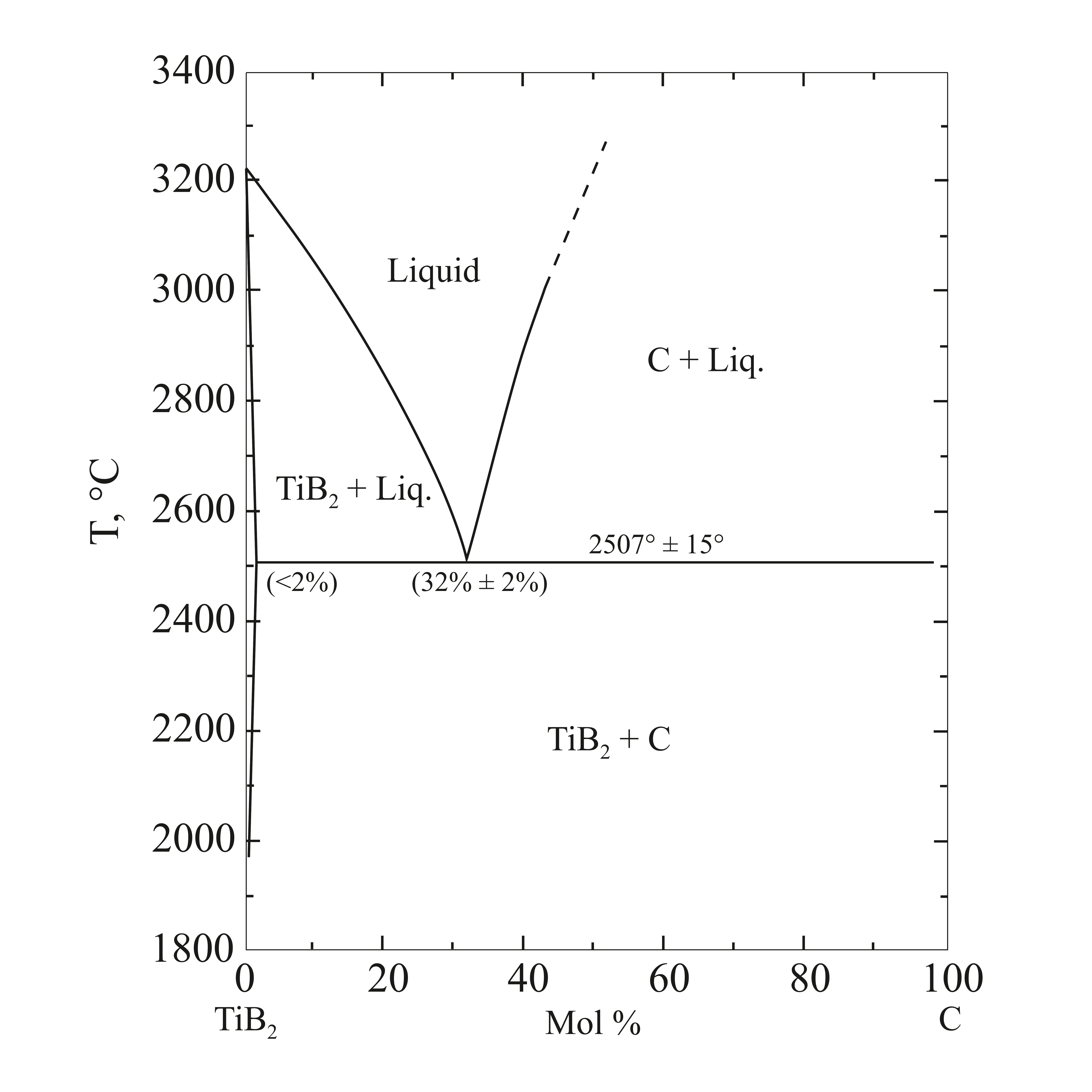}}
  }

  \caption{ (a) Pseudo-binary sections under consideration are colored as follows: orange for Line 1 and green for Line 2; (b) Pseudo-binary section TiB$_2$-TiC$_{0.95}$ herein Line 1; (c)Pseudo-binary section TiB$_2$-C, herein Line 2.
  The Figures are drawn according to the data from~\cite{Huang18_UHTC_Part}.}\label{fig_Book}
\end{figure*}

In this paper, a new approach based on artificial intelligence (AI) methods to determine the position of the eutectic in the pseudo-binary section of a multicomponent ultra-refractory alloy is proposed. The approach involves studying the physical properties of the compound at temperatures above the solidus line, which allows determining the eutectic position with acceptable accuracy even without knowledge of the solid phase crystalline structure. This is crucial when searching for new ultra-refractory alloys. Unlike CALPHAD, the proposed method does not require any experimental data (only information on the elemental composition) and has no tuning parameters. Therefore, it can be considered as a useful tool for preliminary specification of eutectics. As for the comparison with CALPHAD, this task is certainly important and will be solved in the future.

\section{Possible indicators of eutectic concentration}
\label{Motiv}

From a naive atomic point of view, an eutectic corresponds to a situation where dissimilar atoms, accounting for their concentrations and the characteristics of the interatomic potential, are ''most favorable`` to cooperate in an alloy. In the solid sphere model, we would expect to find a local maximum in the packing density at the eutectic composition. However, in a real system, and in numerical simulations with any realistic potential the size of an atom is not well-defined.

Instead, the physical density $\rho$ serves as a useful indicator when the simulation is carried out in the NPT ensemble (constant number of particles, pressure, and temperature).
In the NPT ensemble the relaxation of  the system is due to the volume change with a fixed total mass. Therefore, it can be expected that the density will be sensitive to optimal packing.

Another indicator is the total energy of the system. A significant short-range order transformation in a narrow concentration range may lead to a local energy extremum.

The third sensitive quantity, that can react to optimal packing, is diffusion coefficient (mean squared displacement per unit time), which is related to the root mean squared displacement (MSD). MSD measures the average distance that a particle travels from its initial position over time. This is a common indicator of random motion, and it should be strongly affected by optimal packaging.

We have also calculated two another quantities --- bulk modulus and viscosity coefficient --- that are expected to be sensitive to the eutectic point.

So, we have chosen five quantities --- density, total energy, diffusion coefficient, bulk modulus and viscosity coefficient as the core indicators for determining the eutectic point. We have also calculated concentration derivatives for density and energy to enhance the desired effect.

Our choice was conditioned by two factors. First, as it will be seen later, these values clearly demonstrate features in the eutectic concentration field. Secondly, they are convenient to obtain within the computational scheme described below.
Of course, there are other characteristics of the system that are sensitive to the eutectic point, such as thermal conductivity. The numerical calculation of these quantities will be the subject of further studies.

\begin{figure}[h]
  \centering
  \includegraphics[width=0.45\textwidth]{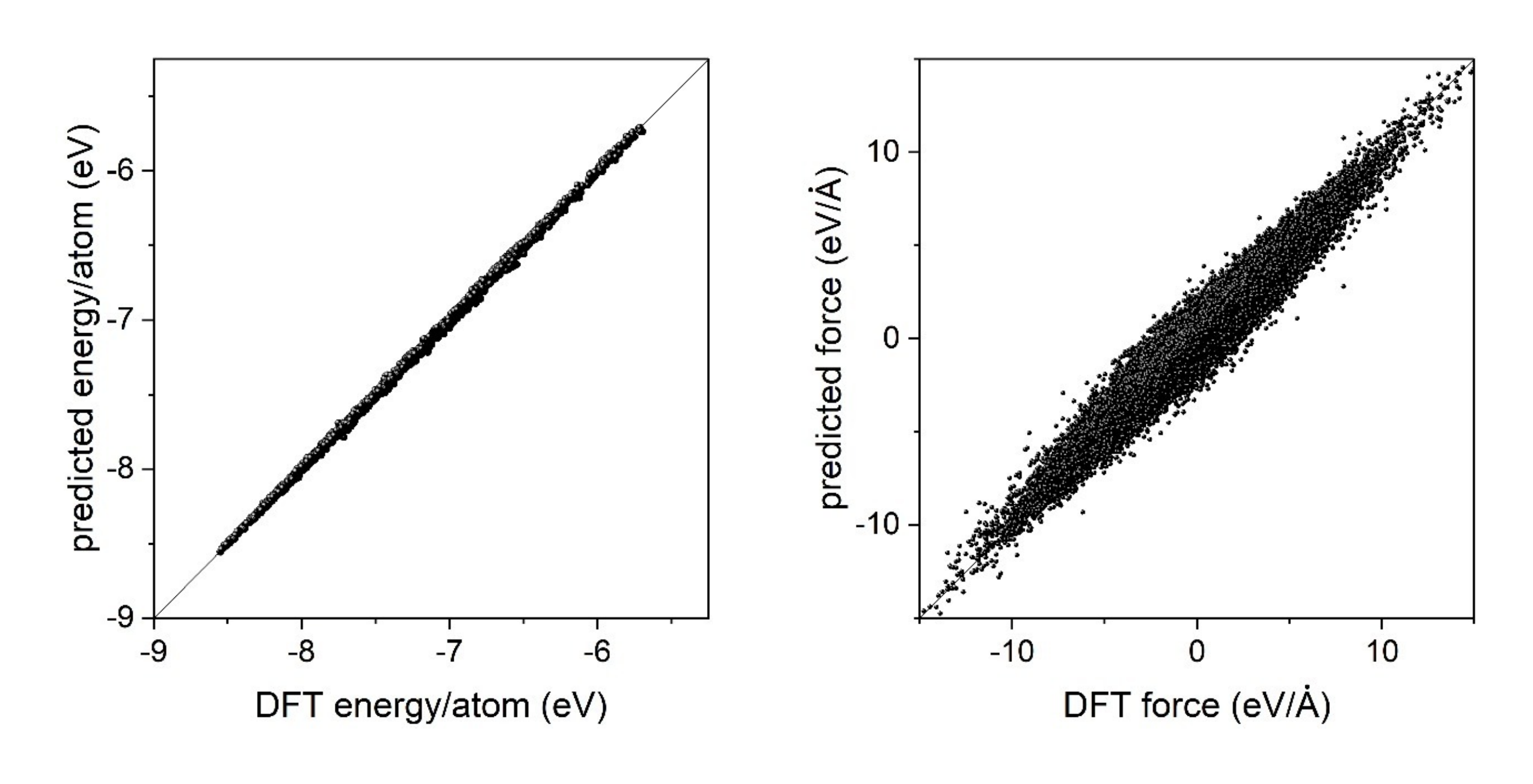}
  \caption{\  Validation of the machine-learning interatomic potential (MLIP) accuracy. The figure reports the model's performance on the combined training and test datasets. When evaluated strictly on the held-out test configurations, the model achieves a root-mean-square error (RMSE) of $14.2$~meV/atom for energy and $0.5$~eV/\AA{} for forces. The reported force error is relatively high, which can be attributed to the complexity of simulating a high-temperature liquid alloy composed of chemically distinct elements (Ti, alongside the inherently incompressible elements boron and carbon), combined with the use of a limited training dataset. Nevertheless, the potential successfully passes a series of validation tests and correctly reproduces key thermodynamic features, such as the eutectic points. This demonstrates that even a modest model can reliably capture complex materials behavior---a central finding of our work. }\label{fig_DFT}
\end{figure}

Note that these effects can be significantly smeared in melts composed of elements with similar properties, where optimal packing can hardly be detected. A similar phenomenon can be observed in alloys composed of atoms with significantly different sizes, where small atoms can easily fill the spaces between larger ones. Fortunately, most of the practically important refractory alloys do not fall into these categories.

Let us also mention, that contribution of structural factors to the eutectic behavior is a rather complex problem, and the detailed discussion was not the essential aim of the work. Here we simply proposed a numerical method to narrow down the search area for the eutectic composition and thereby to simplify the investigation of new ultra-refractive compositions.

\section{Computational scheme}
\label{Subj}

In this paper, we use numerical simulation to demonstrate the peculiarities of the liquid melt properties in the eutectic region. As a test case, we take the three-component Ti-B-C system, which is the most experimentally studied three-component ultra-refractory system~\cite{Banerj13_AM,Huang18_UHTC_Part}.

Two lines (pseudo-binary sections) are selected on the Gibbs-Roseboom triangle(Fig.~\ref{fig_Book}a).

The first one (hereafter referred to as Line 1, Fig.~\ref{fig_Book}b) is TiB$_2$-TiC (more precisely, TiC$_{0.95}$, but boundary effects are not important in this work). On this line, the eutectic concentration is experimentally determined as $57\pm 2\%$~\cite{Huang18_UHTC_Part}.

The second line (herafter referred to as Line 2, Fig.~\ref{fig_Book}c) --- TiB$_2$-C. Here, the eutectic concentration is experimentally determined as $32\pm 2\%$~\cite{Huang18_UHTC_Part}.

The general computational scheme was as follows.

Step 1 (Section~\ref{Sample}): We use the standard VESTA functionality to create a $\beta$ titanium lattice.

Step 2 (Section~\ref{Neuro}): Using DeepMD's meta-deep learning approach, we carefully create a neural network potential based on an accurate quantum mechanical calculations generated by VASP.

Step 3 (Section~\ref{MD}): Molecular dynamics framework poorly describes subtleties of  melting and crystallization processes, so we increase the temperature until the effects of these processes are less significant. We heat the sample from Step 1 to a temperature significantly above the melting point of titanium: around $5000 \ K$. Then, we randomly replace titanium atoms with boron and/or carbon to achieve the desired concentration. After that, we stabilize the sample in N-P-T ensemble at target $T = 2500 \ K$: we run the simulation as long as all the decisive characteristics, such as energy, density, etc. remain constant (slightly varying).

Once the sample has been stabilized, we cool it below the melting point to avoid the subtleties associated with the crystallization process and to enhance the effects associated with the eutectic point. We also performed test calculations at temperatures above the solidus. Desired effects are observed there, but they are weaker and less clear, which reduces the accuracy in determining the eutectic point.
It is important to note that all the results weakly depend on temperature changes. Nevertheless calculating all characteristics in a wide temperature range can be computationally intensive and will be discussed in our future work.

After that, density, total energy, diffusion coefficients, viscosity, and bulk modulus were calculated along both lines in the N-P-T ensemble. The calculations were not performed over the entire concentration range, but over a fairly wide range, which is enough for the eutectic search. There were $16 000$ particles in the simulation. Note that qualitatively and semiquantitatively, all the results were reproduced at smaller and larger values of  $N$, and further increase of particles' numbers does not significantly affect the outcomes, but increase the computational time dramatically.

\section{Numerical sample preparation}
\label{Sample}

Initially, the VESTA~\cite{Momma11_JAC} (see e.g. recent applications
\cite{Celtek25_JCP,Della25_JCP,Fukui25_JCP,Vrba25_JCP}) program produced a digital image of a $\beta$ titanium crystal. Then it was melted (by heating to a temperature of $T > 4000 \ K$). Then, in the ''liquid`` sample, using the Atomsk program~\cite{Hirel15_CPC} (see e.g. recent applications \cite{Masuda24_PRB,Singla24_PRB,Zhang24_PRB}, the required number of randomly selected Ti atoms in the desired proportion was replaced with B and C atoms (in accordance with the total number of atoms, and the target concentrations). After a sufficiently long relaxation, the temperature dropped to the operating one and final calculations were performed.

We done the final calculations at $T = 2500 \ K$, that is, below the solidus line. Crystallization in molecular dynamics, if it occurs, then takes place in times significantly exceeding the calculation time. In other words, the results below were actually obtained for a supercooled liquid. We also performed calculations at temperatures above the solidus. Desired effects are observed there, but they are weaker and less clear, which reduces the accuracy in determining the eutectic point.

\section{Structure and training of  the neural network}
\label{Neuro}

\begin{figure*}[thb]
  \centering
  \includegraphics[width=0.95\textwidth]{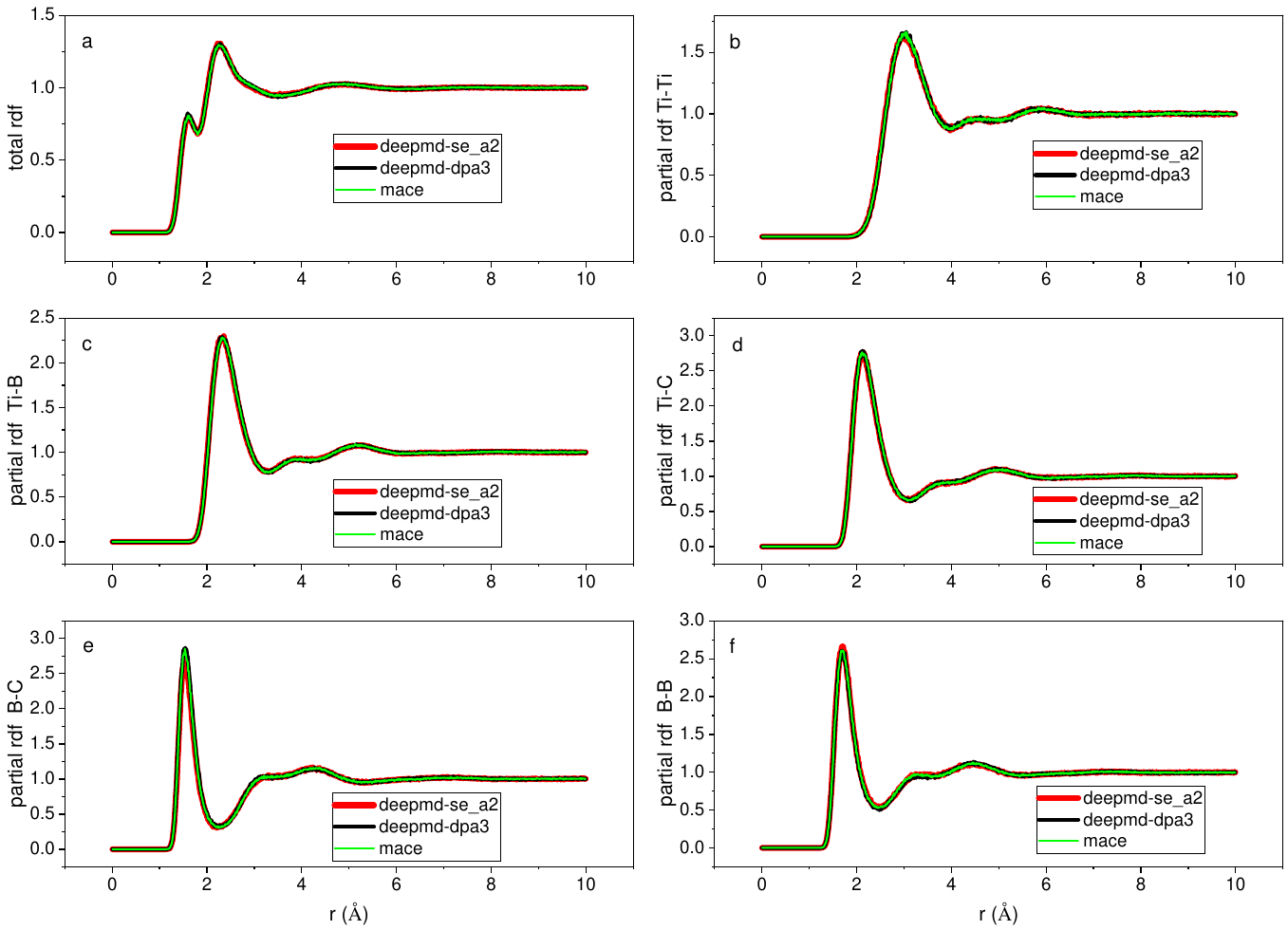}
  \caption{\  Accuracy assessment of the DeepMD-se-a2 MLIP via radial distribution function (RDF) benchmark.
Total and partial radial distribution functions for a liquid $\mathrm{Ti_{40}B_{40}C_{20}}$ alloy
(obtained as a 1:1 molar mixture of TiB$_2$ and TiC) at $T = 4000$~K.
The computationally efficient descriptor-based DeepMD-se-a2 potential is compared against two
high-accuracy graph-based foundation models, DeepMD-DPA3 (DPA-3.1-3M) and MACE (mace-mp-0b3-medium),
both fine-tuned on the DeepMD-se-a2 training dataset.
The close agreement between all calculated RDFs confirms the accuracy of the optimized DeepMD-se-a2 MLIP,
despite its relatively high reported force error of $0.5$~eV/\AA.
Additional benchmarks are presented in Fig.~\ref{fig_VACF}. }\label{fig_rdf_dpa3f}
\end{figure*}
\begin{figure*}[htb]
  \centering
  \includegraphics[width=0.85\textwidth]{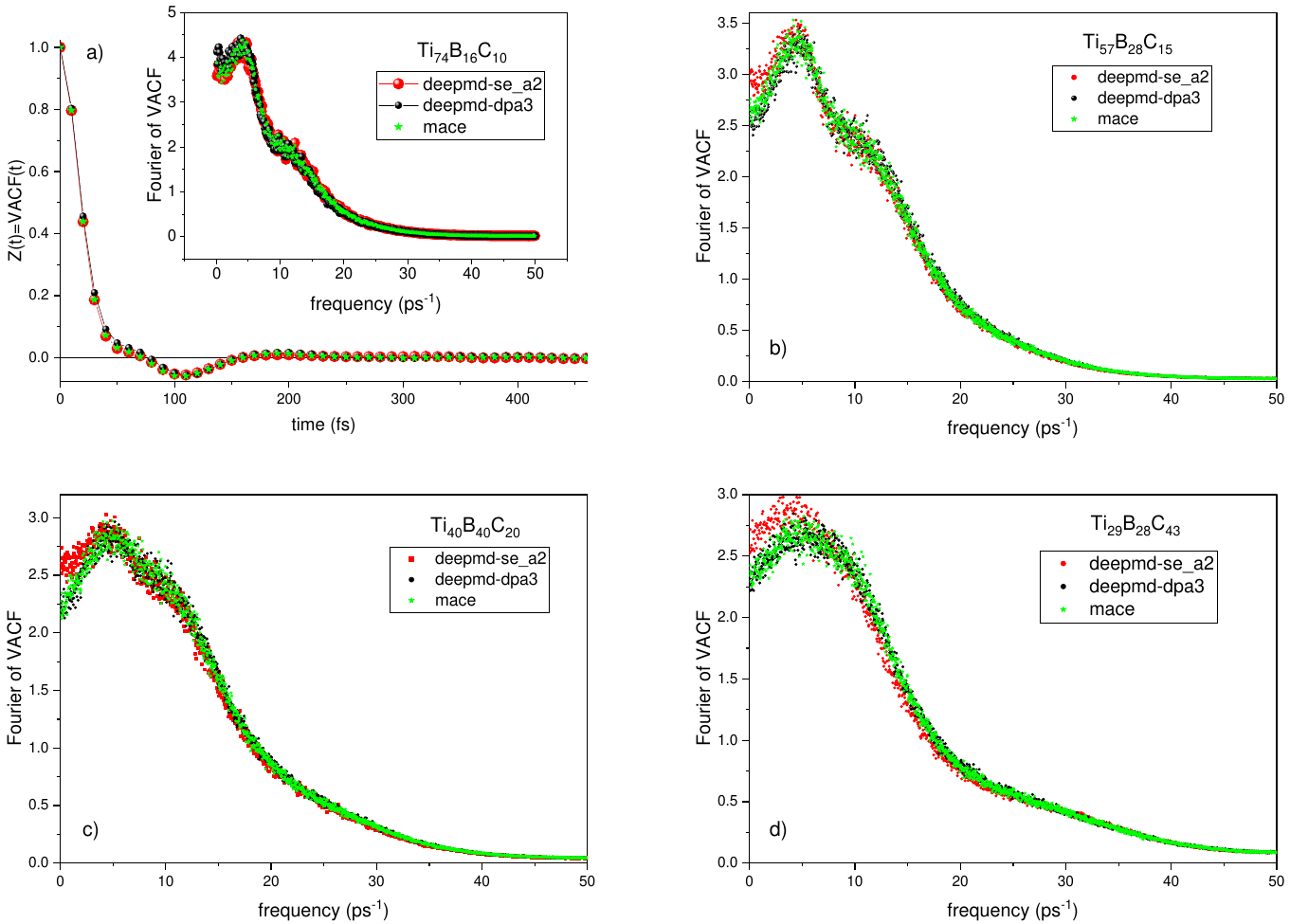}
  \caption{\  Accuracy assessment of the DeepMD-se-a2 MLIP via velocity autocorrelation function (VACF) benchmark. The computationally efficient descriptor-based DeepMD-se-a2 potential is compared against two different high-accuracy graph-based MLIP foundation models DeepMD-DPA3 (DPA-3.1-3M) and MACE (mace-mp-0b3-medium), both fine-tuned on the DeepMD-se-a2 training-test dataset. (a) VACF for a liquid Ti$_{74}$B$_{16}$C$_{10}$ alloy at $T = 4000~\text{K}$, computed in the NVT ensemble. The inset shows the Fourier transform of the VACF; the low-frequency noise is attributed to the finite simulation trajectory length (10~ps). For panels (b)--(d), computations were performed in the NPT ensemble. The larger discrepancy between MLIP and GMLIP at low frequencies stems from enhanced pressure fluctuations in the NPT-MLIP simulations at high temperature.}\label{fig_VACF}
\end{figure*}

The key feature of the computational procedure is the use of the machine-learning interatomic potential (MLIP). A detailed consideration of MLIP is not the subject of this work, so here we refer only to a few recent articles
~\cite{Derin19_AM,Chtche23_JCP,Khazie24_JCP,Goswam25_JCP,Hanser25_JCP,Tang25_JCP,Matsut24_JCP,Matsut24_JCP,Li24_PRB,Nguyen24_PRB,Li24_CMS} and the book~\cite{Mortaz23_Book}

We used the DeePMD~\cite{Wang18_CPC} neural network, see e.g. recent applications \cite{Dupuy24_JCP,He25_JCP,Infuso25_JCP,Tang25_JCP,Yu25_JCP}. The neural network was trained on samples with the appropriate composition using the first-principle quantum mechanical package VASP \cite{Hafner97_Book_VASP}.

For the atomic-scale modeling of Ti-B-C systems, we developed a machine learning interatomic potential (MLIP) trained on first-principles density functional theory (DFT) data. The training dataset comprised unit cell structures, atomic coordinates, interatomic forces, virial stresses, and potential energies. The potential was constructed using the DeePMD framework~\cite{Wang18_CPC} with ``se-a''-type descriptors~\cite{Zhang18_ANIPS}, employing a cutoff radius of 7~\AA{} and neural network architectures consisting of three hidden layers (25, 50, and 100 angular neurons) complemented by a 20-neuron embedding layer. The fitting network featured three hidden layers with 255 neurons each. The optimized potential achieved root mean square errors (RMSE) of 14.2~meV/atom for energy predictions and 0.5~eV/\AA{} for forces (at test configurations), as demonstrated in Fig.~\ref{fig_DFT}. {\  The training procedure comprised one million ($10^6$) steps with a batch size of 4.}

\subsubsection{Active learning}

To generate initial configurations at each DPGEN iteration (Deep Potential GENerator~\cite{Zhang2020}), we employed the Dirichlet-Rescale (DRS) algorithm~\cite{DRS_1,DRS_2}, which produces uniformly distributed random vectors constrained to unit summation. This approach enabled systematic sampling across the entire concentration space. For each DPGEN step, we created 12 random (Ti, B, C) compositions, initialized as FCC crystals that were subsequently melted at 5000~K using the DeePMD potential from the previous iteration. To prevent unphysical atomic aggregation during melting, we incorporated a hard-sphere repulsive potential with radii parameters determined from radial distribution function analysis. The resulting equilibrated configurations served as initial states for classical molecular dynamics simulations.

The DPGEN algorithm operated via an iterative active learning cycle. At each iteration, four MLIP replicas---initialized with different random seeds---were automatically trained on the accumulated database. These potentials were then used to perform ensemble molecular dynamics simulations, generating new configurations for database expansion. Configurations exhibiting the highest uncertainty in interatomic forces, as quantified by the standard deviation across the ensemble, were selected for first-principles calculations and subsequent inclusion in the training database. This cycle continued until active learning convergence was achieved. All molecular dynamics simulations were performed on systems of 100--150 atoms under periodic boundary conditions, using a time step of 0.1~fs and trajectory lengths of 25,000 steps (2.5~ps). The liquid alloy was simulated during both heating (3000--5000~K) and cooling (5000--2500~K) cycles using a Nosé-Hoover thermostat, with applied pressures ranging from 0 to 50~GPa. Atomic compositions were systematically varied using the DRS algorithm throughout this process. Note, that rather small cell is used only for MLIP training, the final calculations (see Section \ref{MD}) involve more than 15,000 particles.

{\

The training of the DeePMD model employs a multi-term loss function that simultaneously minimizes errors in energy ($\Delta E$), interatomic forces ($\Delta \mathbf{F}$), and virial tensor components ($\Delta \xi$)~\cite{Wang18_CPC,Zeng2023,Zeng2025}.
The composite loss function is defined as
\[
\mathcal{L}(t) = p_E(t) \, \|\Delta E\|^2 + p_F(t) \, \|\Delta \mathbf{F}\|^2 + p_\xi(t) \, \|\Delta \xi\|^2,
\]
where $p_E(t)$, $p_F(t)$, and $p_\xi(t)$ are time-dependent prefactors that evolve during training from initial (\texttt{start\_pref\_*}) to final (\texttt{limit\_pref\_*}) values.

In our training setup, these prefactors were scheduled as follows:
\begin{itemize}
    \item For energy: \texttt{start\_pref\_e = 0.02} $\rightarrow$ \texttt{limit\_pref\_e = 1.0}
    \item For forces: \texttt{start\_pref\_f = 1000} $\rightarrow$ \texttt{limit\_pref\_f = 1.0}
    \item For virial: \texttt{start\_pref\_v = 0.02} $\rightarrow$ \texttt{limit\_pref\_v = 1.0}
\end{itemize}
This scheduling forces the optimization to prioritize force accuracy in the initial stages (\texttt{start\_pref\_f $\gg$ start\_pref\_e}), which is critical for the stability of training and leverages the statistical abundance of force data ($3N$ components per configuration versus a single energy value). While the $\Delta E$ term ensures the thermodynamic consistency of the potential energy surface, the dominant $\Delta \mathbf{F}$ term guides the early optimization. The $\Delta \xi$ term, initialized with a low prefactor, introduces constraints relevant for simulations under non-zero pressure.

By the end of the optimization, all prefactors converge to the same order of magnitude (\texttt{limit\_pref\_* $\sim$ 1}), ensuring a balanced contribution from each term in the final model.
}

Finally, DPGEN active learning converged after 20 iterations, leading to $~$4500 configurations in the training database and about 500 structures for tests.

{\  First-principles density functional theory (DFT) calculations were performed using the Vienna Ab Initio Simulation Package (VASP)~\cite{Hafner97_Book_VASP}.
A plane-wave basis set with a kinetic energy cutoff of 600~eV was employed.
We used the generalized gradient approximation (GGA) with the Perdew--Burke--Ernzerhof (PBE) exchange--correlation functional~\cite{Kresse1999}.
Projector augmented-wave (PAW) pseudopotentials were utilized, which explicitly treat not only valence electrons but also semicore states.
The Brillouin zone was sampled using a $\Gamma$-centered $k$-point mesh generated via the K-Point Grid Server~\cite{Wisesa2016,Wang2021} with a density of approximately 3000 $k$-points per reciprocal atom (\texttt{MINDISTANCE} = 36).
Electronic convergence was achieved with an energy threshold of $10^{-6}$~eV (\texttt{EDIFF} = $10^{-6}$).}

\subsubsection{MLIP verification}

{\
To validate the accuracy and reliability of our newly developed Deep Potential Molecular Dynamics (DeePMD) potential for molten Ti–B–C systems, we conducted a series of comprehensive benchmark tests.
While graph neural network (GNN)-based interatomic potentials generally provide higher accuracy than simpler, embedding-based architectures like DeePMD-se-a2, they incur a substantially higher computational cost.
For comparison, we selected two universal foundation models: DeePMD-DPA3 (DPA-3.1-3M) and MACE (mace-mp-0b3-medium).
Both models were pretrained on extensive databases, including the Materials Project (MP)~\cite{Jain2013,Zhang2025,Batatia2025}, and were subsequently fine-tuned on our custom \textit{ab initio} dataset generated via VASP calculations and the DP-GEN active learning cycle.
The fine‑tuning protocol used 100,000 steps with a batch size of 4 for DeePMD-DPA3 and 200 steps for MACE, reflecting the different training dynamics and convergence behavior of the two architectures.

Given the higher computational demand of GNN-based potentials, all benchmarks were performed on relatively small systems of 500 atoms.
The fine-tuned GNN potentials achieved the following root-mean-square errors (RMSE) on test configurations:
DeePMD-DPA3 reached 4.6~meV/atom for energy and 0.23~eV/\AA{} for forces,
while MACE achieved 4.2~meV/atom for energy and 0.25~eV/\AA{} for forces.

Our validation approach is based on a rigorous, multi-model consistency test.
We compute a set of critical liquid-state properties (e.g., radial distribution functions, velocity autocorrelation functions) via molecular dynamics simulations using three interatomic potentials with different foundational architectures:
(1) the computationally efficient but less accurate DeepMD-se-a2 potential, and
(2) two high-accuracy graph-based potentials (DeepMD-DPA3 and MACE), which themselves employ fundamentally different internal neural network structures and message-passing schemes.

This design serves as a powerful sensitivity test for the quality of the underlying training data.
Graph neural networks, with their higher expressive power, are more sensitive to the quality, coverage, and consistency of the training dataset.
If the dataset were insufficient, incomplete, or contained systematic biases, these architecturally distinct models would be likely to produce significantly different predictions.
Conversely, the convergence of results obtained from all three models provides strong, architecture-agnostic evidence that the training dataset is representative and that the simpler DeepMD-se-a2 potential has successfully captured the essential physics of the molten Ti-B-C system, making it a reliable and efficient tool for large-scale simulations.}

Figures~\ref{fig_rdf_dpa3f}--\ref{fig_VACF} illustrate radial distribution functions (RDFs) and the velocity autocorrelation functions (VACFs) for the melt at a representative composition, simulated using two GNN-based machine learning interatomic potentials and the se-a2 MLIP. We verified the consistency of these methods across a broader range of compositions (not shown), observing no significant discrepancies. The close agreement with \textit{ab initio} reference data confirms that the se-a2 MLIP achieves sufficient accuracy for this class of systems in a liquid state.

\begin{figure}[h]
  \centering
  \includegraphics[width=0.4\textwidth]{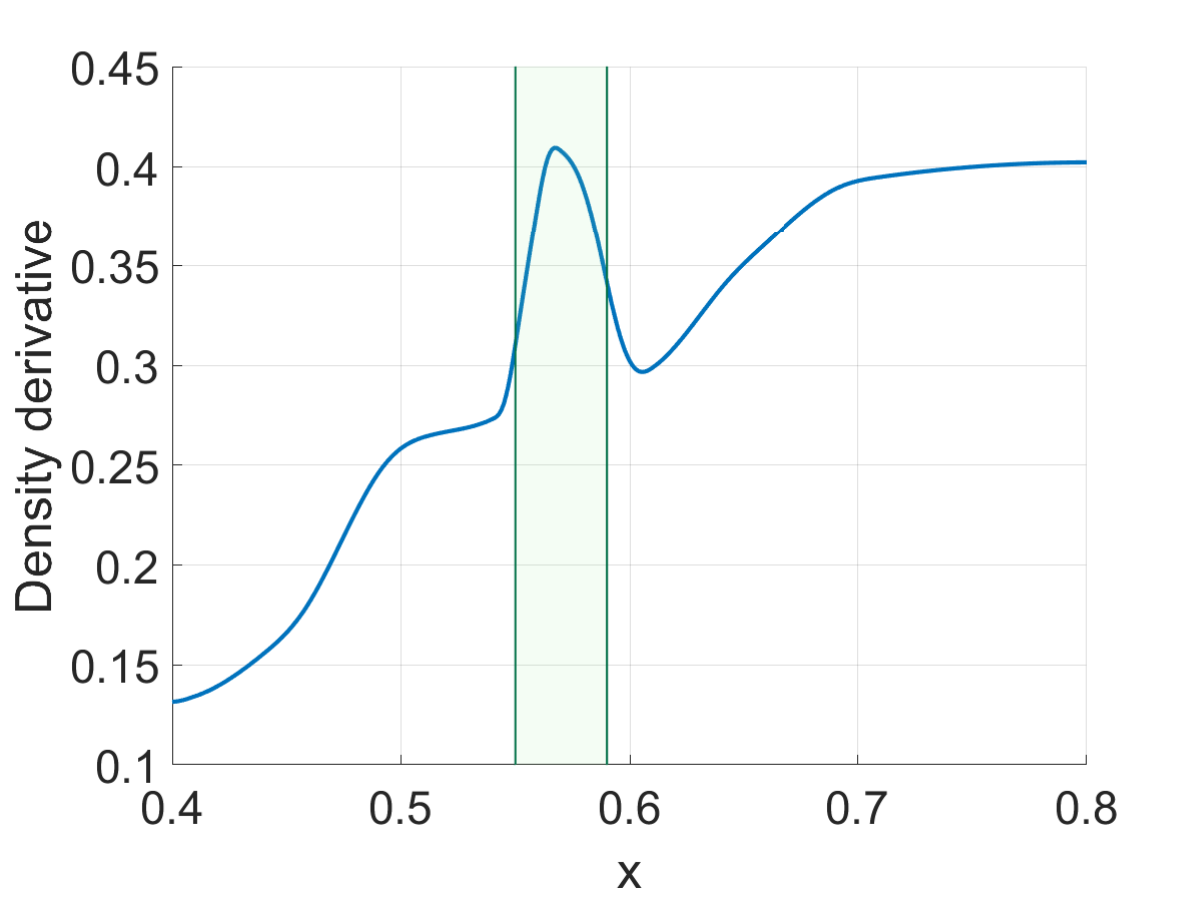}
  \caption{The derivative of density by concentration as the function of concentration in Line 1 is shown (density in $g/cm^3$ units, common in LAMMPS {\itshape metal} style). The experimentally estimated eutectic region~\cite{Huang18_UHTC_Part} is shaded
  ($57 \pm 2 \ Mol \%$). Against the background of a monotonous increase in density (and its derivative) in the non-eutectic region, a local maximum is observed in the eutectic region.}\label{fig_Line1_Dens}
\end{figure}

\begin{figure}[h]
  \centering
  \includegraphics[width=0.4\textwidth]{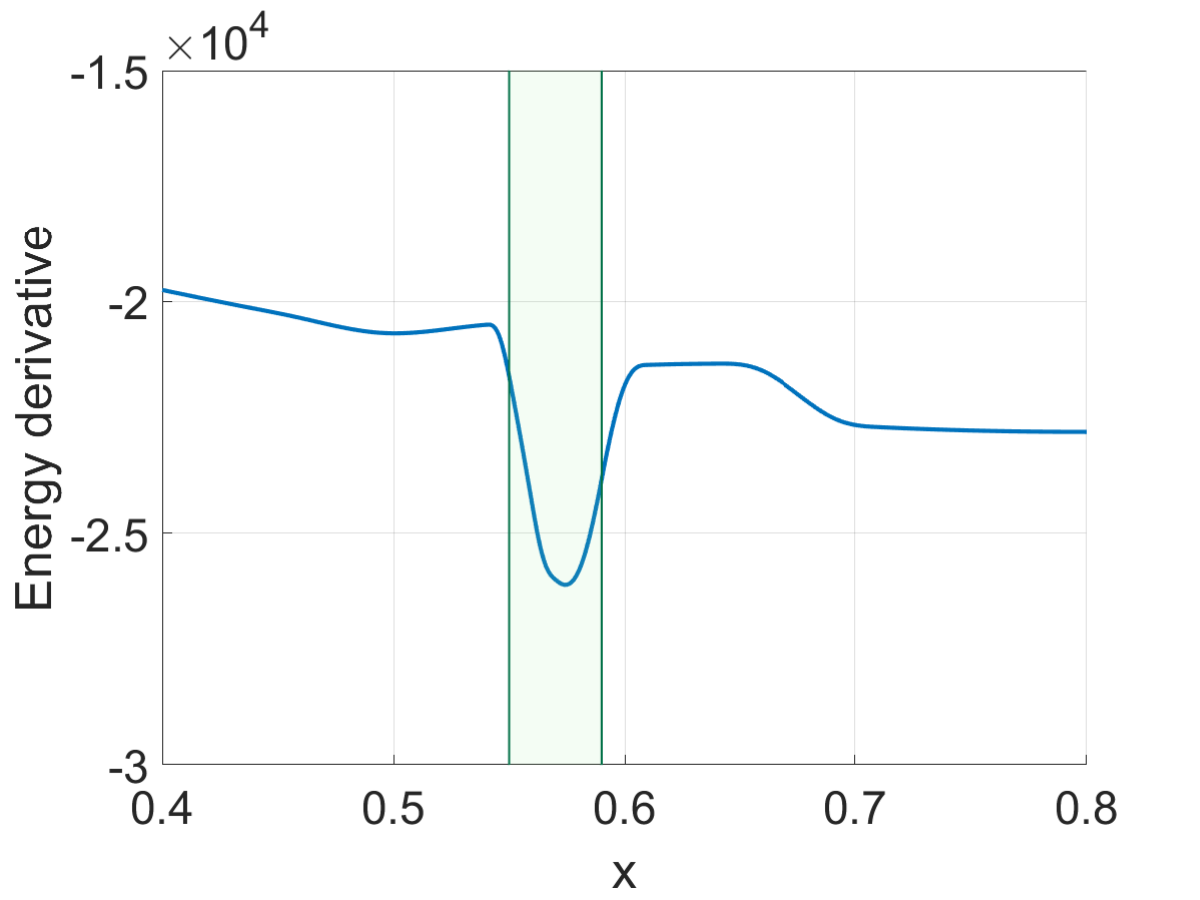}
  \caption{The derivative of the total energy by concentration as the function of concentration in Line 1 is shown (energy in $eV$ units, common in LAMMPS {\itshape metal} style). The experimentally estimated eutectic region~\cite{Huang18_UHTC_Part} is shaded ($57 \pm 2 \ Mol \%$). As in the previous Figure, a local feature occurs in the eutectic region.}\label{fig_Line1_Ener}
\end{figure}

To further test the accuracy of the neural network's potential for interaction between atoms, a number of additional calculations were performed. MLIP was trained on a database of \textit{ab-initio} calculations, which contained mostly disordered configurations, as studies of melts were assumed. The most rigorous test of MLIP consists in verifying MLIP's predictions for crystalline phases that are stable at low temperatures, which were not present in the training database.

TiC [aflow:f658528958f9b1a3] and TiB2 [aflow:207fb1b3378121cd], compounds located on the convex hull of Ti-B-C, according to the AFLOW database~\cite{Curtarolo2012,Curtarolo2012a,Calderon2015} (\url{https://www.aflowlib.org/}), as well as the metastable TiBC compound (mp-1232377 from the Materials Project database~\cite{Jain2013,Horton2025}, \url{https://next-gen.materialsproject.org/}) were selected for testing MLIP.

To further complicate the MLIP test, the crystal structure of TiC and TiB2 was guessed with the evolutionary algorithm USPEX~\cite{Lyakho13_CPC} at temperature T=0, where MLIP was used as an interaction potential, and the LAMMPS atomistic modeling package \cite{Thomps22_CPC} was used to minimize energy (we used the {\itshape metal} LAMMPS style in all the calculations, that means the particular choice of the units) \cite{Buglak25_JCP,He25_JCP,Mosca25_JCP,Rele25_JCP,Zhang25_JCP}.

TiC has a cubic fcc structure and $Fm3m ~(\# 225)$ space group. According to the ab-initio calculation given in AFLOW, the lattice cell parameters are $a=4.3375 \AA$. MLIP+USPEX calculations gave $a=4.3348 \AA$.

TiB2 has a hexagonal crystal structure of $P6/mmm~(\# 191)$. According to the ab-initio calculation given in AFLOW, the lattice cell parameters are $a=b=3.0259 \AA, c=3.2288 \AA$. MLIP+USPEX calculations gave $a=b=3.0933 \AA$, and $c=3.1238 \AA$.

TiBC, according to the ab-initio calculation given in the Materials Project, has a hexagonal $P6_3  /mmc ~(\# 194)$ structure with $a=b=2.8295 \AA$ and $c=7.0613 \AA$. Relaxation of this structure performed using MLIP and LAMMPS gave the following results: $a=b=2.8265 \AA$ and $c=7.0539 \AA$.

Summing up the results of these rigorous tests, which went far beyond the range of thermodynamic parameters for which MLIP was optimized (we prepared MLIP for melts), we can conclude that its quality is quite high.


\begin{figure*}[tph]
  \centering
  \mbox{
    \subfigure[Diffusion coefficient] {\includegraphics[width=0.33\textwidth]{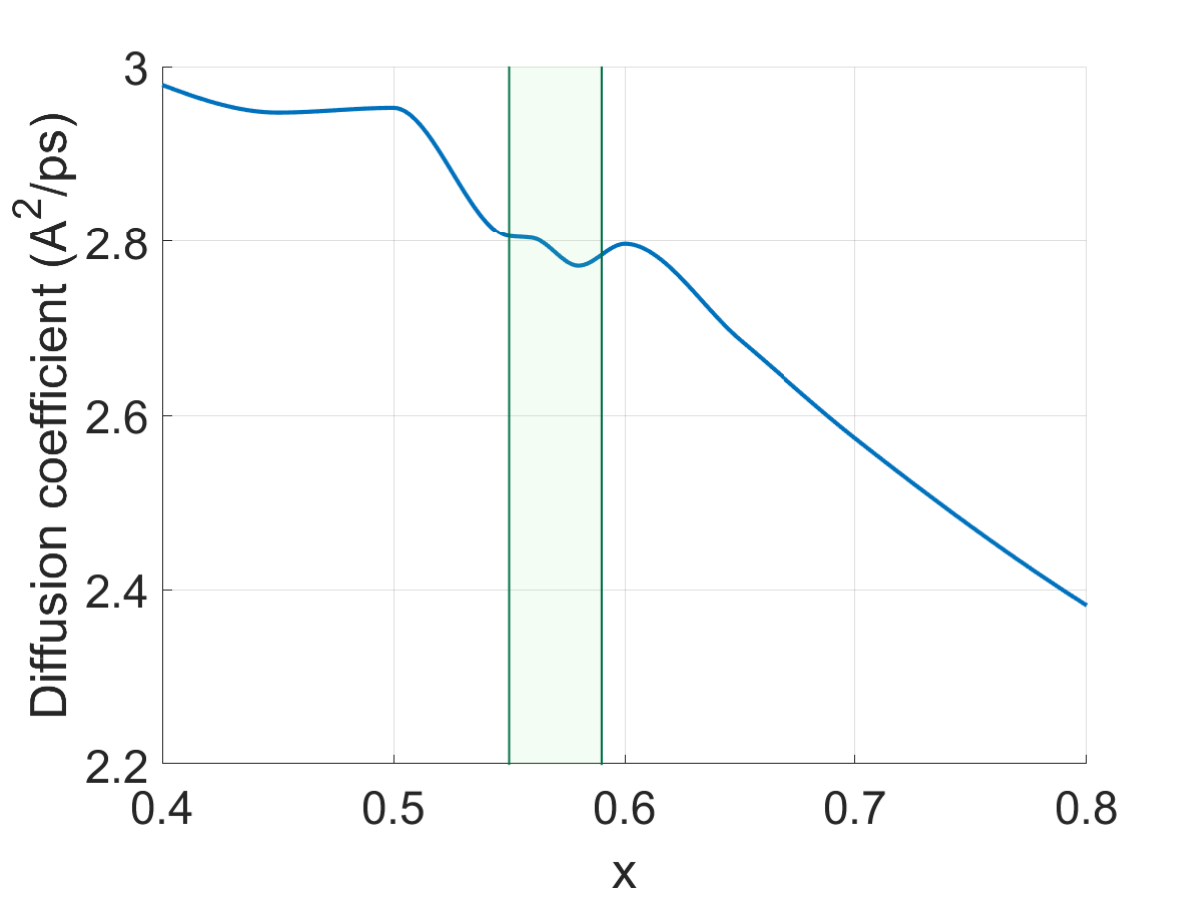}}
    \subfigure[Bulk modulus] {\includegraphics[width=0.33\textwidth]{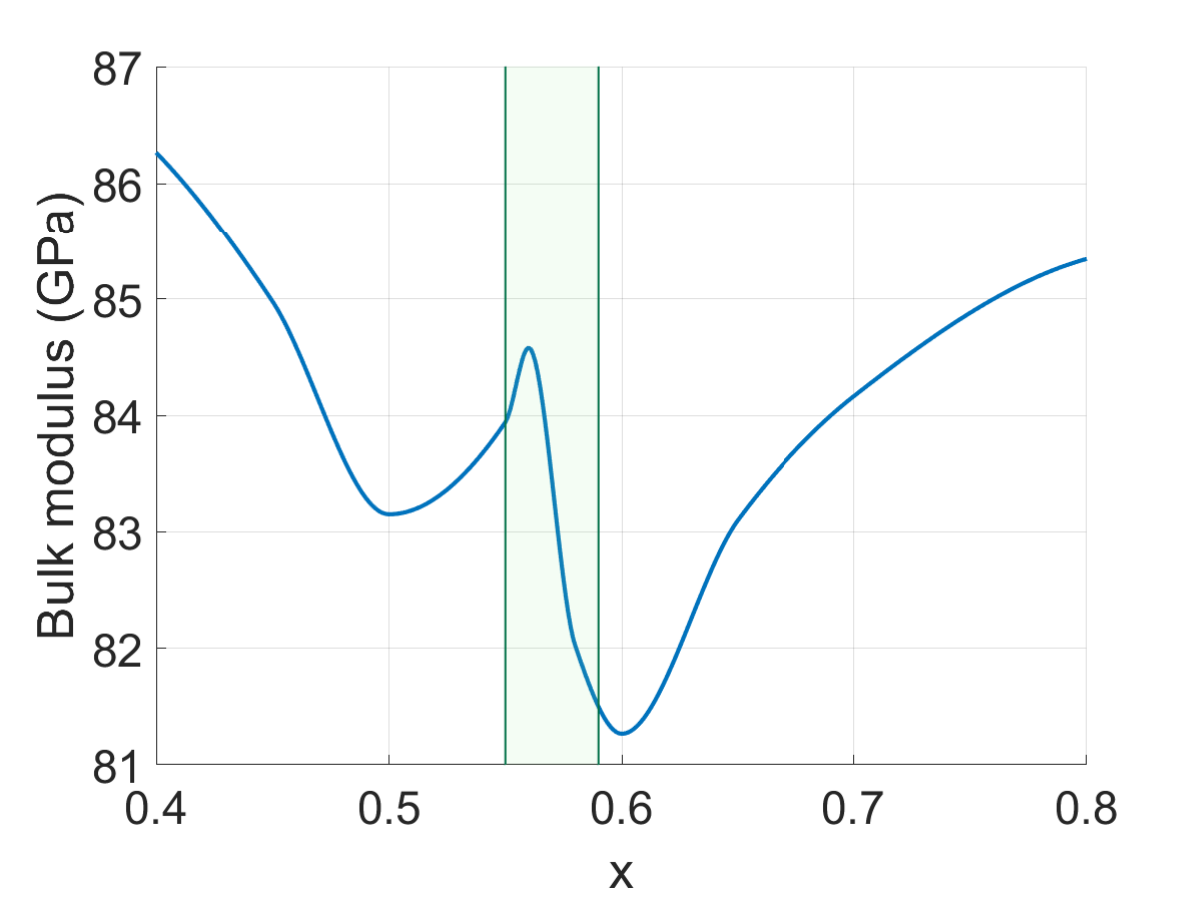}}
    \subfigure[Viscosity coefficient] {\includegraphics[width=0.33\textwidth]{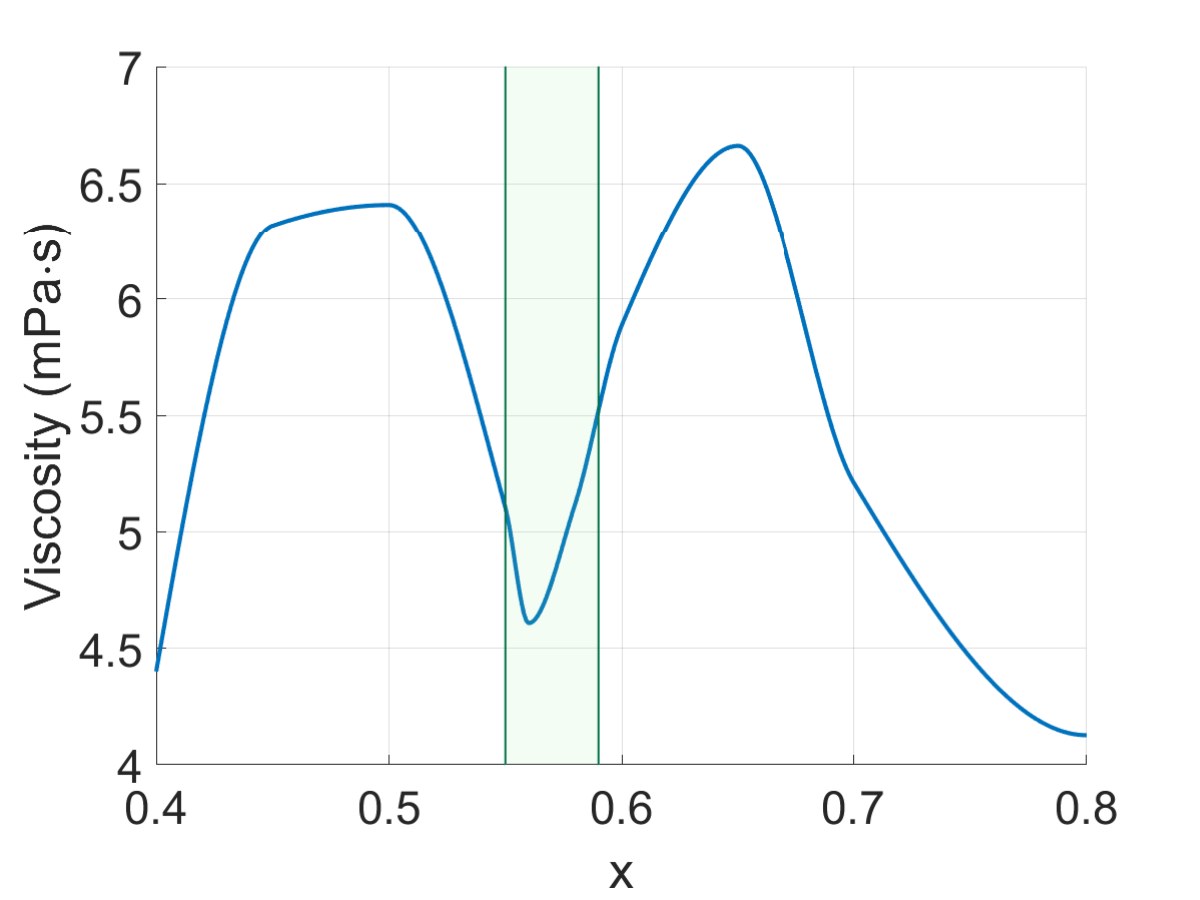}}
  }

  \caption{Dependence on concentration in Line 1. The experimentally estimated eutectic region~\cite{Huang18_UHTC_Part} is shaded ($57 \pm 2 \ Mol \%$). All three quantities exhibit anomalies in the eutectic region, perceptably visible for bulk modulus and viscosity.}\label{fig_Line1_other}
\end{figure*}

The MLIP machine learning database was created using DPGEN active learning algorithms. These algorithms worked by automatically training four MLIP replicas on a database obtained during the early stages of active learning. They also launched an ensemble of classical molecular dynamics (CMD) with this MLIP, selected the least accurate configurations of atoms based on the standard deviation of the interatomic forces averaged over the ensemble. The selected configurations were then sent to the first-principle calculation, and then added to the training database. The cycle was repeated until active learning converged. For CMD modeling of the melt, a cell with 100-150 atoms under periodic boundary conditions was selected. The time step was 0.1 fs, and the length of the trajectory was typically 25,000 steps. The alloy was simulated when heating from 100 Kelvin to 1,500 Kelvin (or cooling from 1,000 Kelvin) using a Nosé-Hoover thermostat in a pressure range of 0 to 50 GPa. The chemical composition of the atoms in the cell was determined by a random algorithm. Note, that rather small cell is used only for MLIP training, the final calculations (see Section \ref{MD}) involve more than 15,000 particles.

The first-principle density functional calculations were carried out using the VASP~\cite{Hafner97_Book_VASP} software package. A plane wave basis with a cut-off energy of 600 eV was used. PAW pseudopotentials were employed, which take into account not only the valence electrons, but also the half-cores.

\begin{figure}[tbh]
  \centering
  \includegraphics[width=0.4\textwidth]{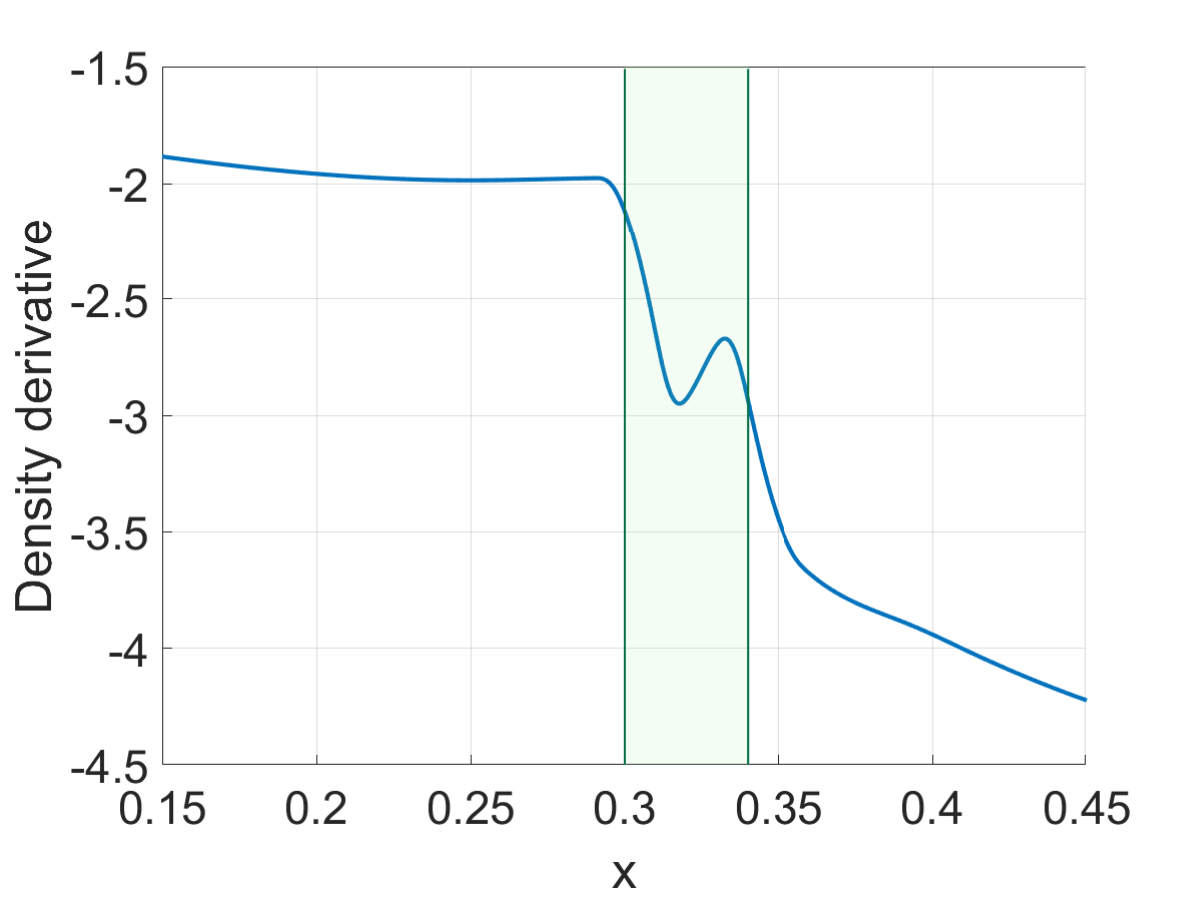}
  \caption{The derivative of density by concentration as the function of concentration in Line 2 is shown (density in $g/cm^3$ units, common in LAMMPS {\itshape metal} style). The experimentally estimated eutectic region~\cite{Huang18_UHTC_Part} is shaded ($32 \pm 2 \ Mol \%$). Against the background of a monotonous reduction in density (and its derivative) in the non-eutectic region, a local feature is observed in the eutectic region.}\label{fig_Line2_Dens}
\end{figure}

\begin{figure}[h]
  \centering
  \includegraphics[width=0.4\textwidth]{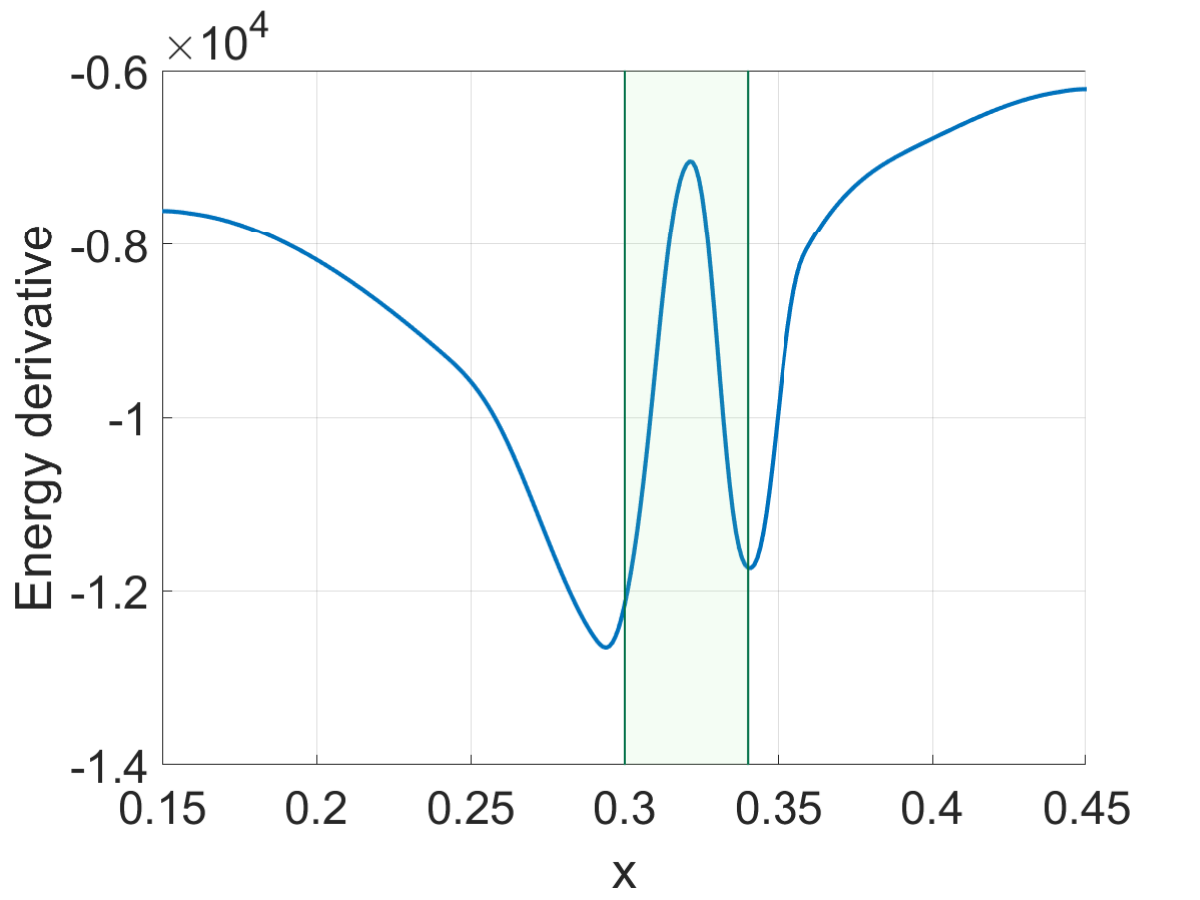}
  \caption{The derivative of the total energy by concentration as the function of concentration in Line 2 is shown (energy in $eV$ units, common in LAMMPS {\itshape metal} style). The experimentally estimated eutectic region~\cite{Huang18_UHTC_Part} is shaded ($32 \pm 2 \ Mol \%$). As in the previous Figure, a local feature occurs in the eutectic region.}\label{fig_Line2_Ener}
\end{figure}

\begin{figure*}[tph]
  \centering
  \mbox{
    \subfigure[Diffusion coefficient] {\includegraphics[width=0.33\textwidth]{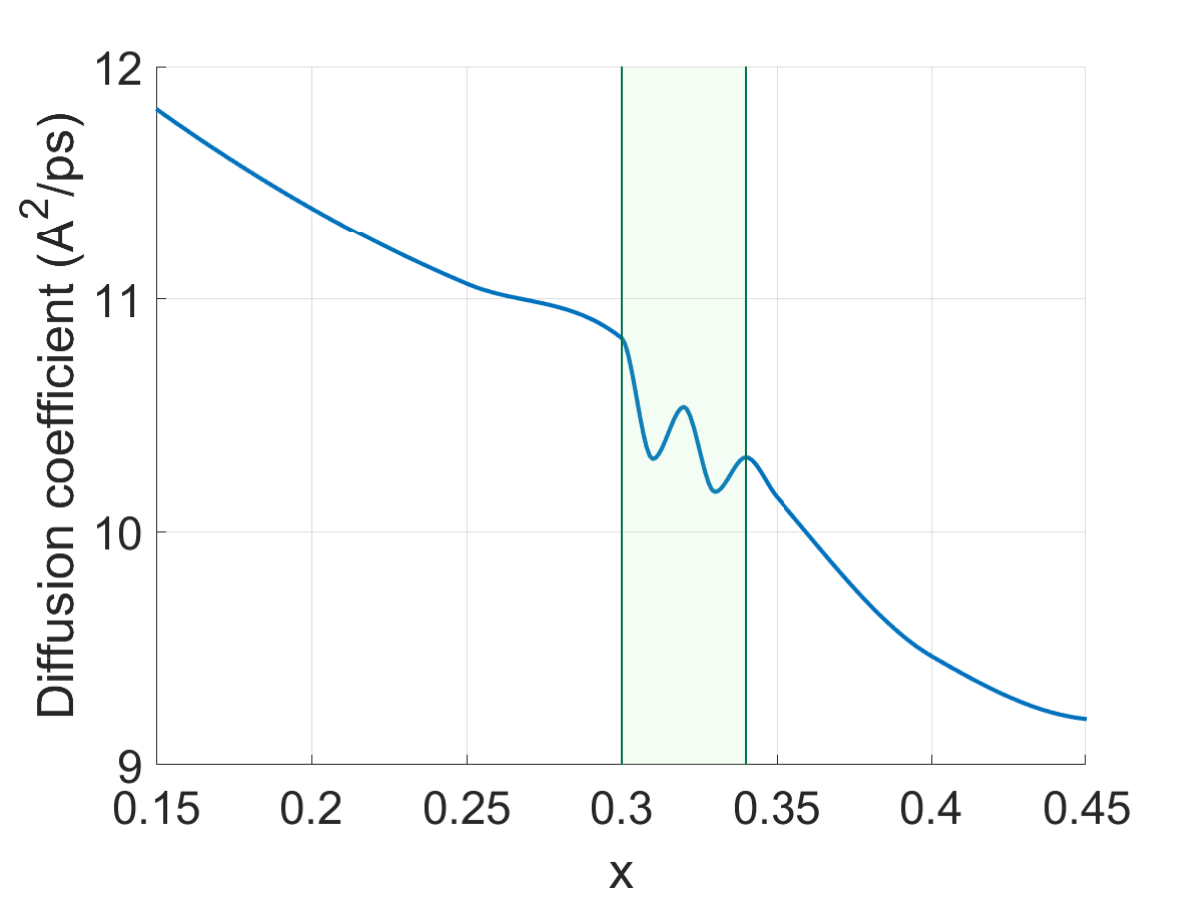}}
    \subfigure[Bulk modulus] {\includegraphics[width=0.33\textwidth]{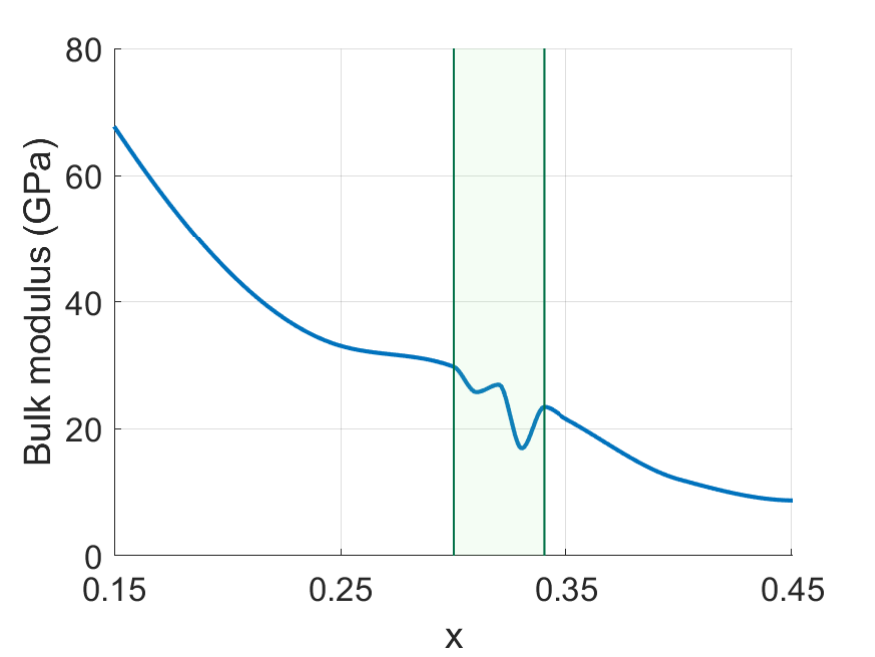}}
    \subfigure[Viscosity coefficient] {\includegraphics[width=0.33\textwidth]{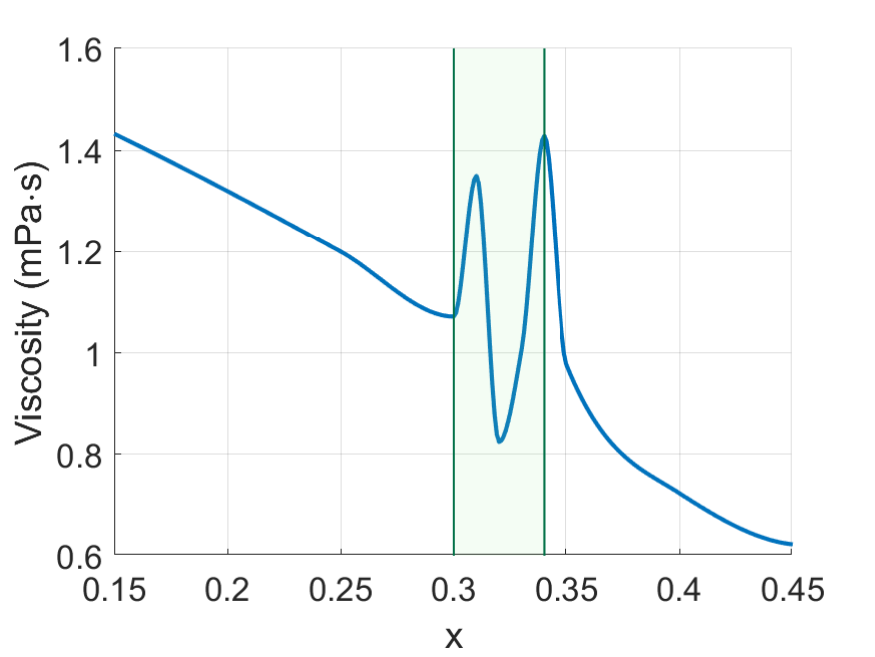}}
  }
  \caption{Dependence on concentration in Line 2. The experimentally estimated eutectic region~\cite{Huang18_UHTC_Part} is shaded ($32 \pm 2 \ Mol \%$). All three quantities exhibit anomalies in the eutectic region.}\label{fig_Line2_other}
\end{figure*}

\section{Molecular dynamics}
\label{MD}

The final calculations were carried out using the molecular dynamics method with the LAMMPS package \cite{Thomps22_CPC}. Molecular dynamics method allows to compute all the basic thermodynamic parameters of a system as well as particle dynamics characteristics such as mean squared displacement, viscosity coefficient, and speed of sound.

As it was mentioned above density, total energy (and their concentration derivatives), diffusion coefficient, bulk modulus and viscosity coefficient were calculated along both lines in the N-P-T ensemble.
The calculations were not carried out over the entire concentration range, but rather over a fairly wide range, that was sufficient for determining the eutectic point. The temperature was set at a level below the solidus line, so we analysed the supercooled melt (to confirm the results we have also recalculated some of them at the temperature above the liquidus line in the studied concentration range).
The pressure was atmospheric.
The number of particles in the simulation $N = 16 000$.
Further increases in $N$ did not lead to any significant changes in the results. Semiquantitatively, all the results were reproduced even at $N\sim 2000, 8000, 10 000$, the results converging with growing $N$.

\section{Results}
\subsection{Line 1}
\label{Line1}

We remind, Line 1 is the pseudo-binary section TiB$_2$ $\rightarrow$ TiC.

\textbf{i.} \textbf{Density.} The most obvious indicator of eutectic concentration is density.
To enhance the effect we have calculated the concentration derivative of the density, it is shown in
Fig.~\ref{fig_Line1_Dens} (hereinafter density and energy are presented in standard Metal style LAMMPS units \url{https://docs.lammps.org/units.html}. The eutectic region is highlighted in color.
In general, there is a monotonic increase in density (and its derivative, shown in Fig.~\ref{fig_Line1_Dens}) over a wide range of concentrations. This is clearly associated with an increase in the proportion of heavier Ti atoms along TiB$_2$ $\rightarrow$ TiC path. However, against this background, a local maximum is observed in the eutectic field. This can be attributed to a denser packing of dissimilar atoms at these mutual concentrations.

\textbf{ii.} \textbf{Energy.} The energy also demonstrates clearly seen feature in the eutectic region. Fig.~\ref{fig_Line1_Ener} shows the concentration derivative of the
the total energy along Line 1 in the N-P-T ensemble (fixed number of particles, fixed pressure and temperature). The eutectic region is highlighted in color.
Apart from the eutectic region the energy derivative is almost monotonic.
But near the eutectic field, its behavior is dramatically different, thus indicating the sought concentration.

\textbf{iii-v.} \textbf{Diffusion coefficient, bulk modulus, viscosity.}
Fig.~\ref{fig_Line1_other} shows three other calculated indicating quantities along Line 1. These are
the diffusion coefficient (mean square displacement per unit time) Fig.~\ref{fig_Line1_other}a, the bulk modulus Fig.~\ref{fig_Line1_other}b, and the coefficient of viscosity Fig.~\ref{fig_Line1_other}c. The eutectic region is highlighted in color.
The situation here is somewhat more peculiar than in the previous two cases.
All three of indicators do demonstrate more or less pronounced features in the vicinity on eutectic concentration.

\textbf{vi.} \textbf{Other characteristics.} It has been mentioned above that the eutectic region  should also be manifested in other properties, such as heat capacity and thermal conductivity.

We believe that the aforementioned results allow to narrow the search for eutectic in the expensive and time-consuming experimental investigations of new ultra-refractory compositions.
Nevertheless, the numerical calculation of other indicating quantities will be the subject of further studies.

\subsection{Line2}
\label{Line2}

Line 2 is the pseudo-binary section TiB$_2$ $\rightarrow$ C.

Fig.~\ref{fig_Line1_Ener} shows the concentration derivative of the
the total energy along Line 1

\textbf{i-ii.} \textbf{Density and energy.}
The the concentration derivative of the density for Line 2 is presented in Fig.~\ref{fig_Line2_Dens}.
The eutectic region is highlighted in color.
In contrast to Fig.~\ref{fig_Line1_Dens}, there is a monotonic increase in density and its derivative, apart from the eutectic region.
This is clearly associated with an decrease in the proportion of heavier Ti atoms along TiB$_2$ $\rightarrow$ C path. And as in Fig.~\ref{fig_Line1_Dens}, against this background, a local feature is observed in the eutectic field.

The concentration derivative of the energy for Line 2 is presented in Fig.~\ref{fig_Line2_Ener}.
The eutectic region is highlighted in color. The clearly visible feature appears near the eutectic concentration.

\textbf{iii-v.} \textbf{Other characteristics.}
The concentration dependencies of the diffusion coefficient, the bulk modulus and the coefficient of viscosity for Line 2 are shown sequentially in Fig.~\ref{fig_Line2_other}.
In all three cases, there is a distinct feature in the eutectic region that stands out against the background of the relatively smooth change in the corresponding values.
But note, that for all the calculated values in Fig.~\ref{fig_Line2_other}, the features in the eutectic region have a more complex structure than those on Line 1. Instead of a simple maximum or minimum, a sinuslike behavior is observed. This is to be attributed to a complex reorganization of the short range order.

\section{Conclusion}
\label{Conclusion}

The numerical simulations were conducted along two lines within the Gibbs-Roseboom triangle for the Ti-B-C system, at a temperature below the solidus of both solidus lines. Employing a supercooled liquid state amplifies the thermodynamic signatures associated with crossing the eutectic region, allowing the calculated characteristics to identify it accurately.

To refine the boundaries of this detected region, the decisive characteristics must be calculated across a wide temperature range to establish their asymptotic behavior — a focus reserved for future work.

The primary aim of this study was not to pinpoint the eutectic coordinates with high precision — a pertinent goal for novel ultra-refractory alloys—but rather to demonstrate the viability of the proposed approach. This explains the choice of the well-characterized Ti-B-C system for validation. The application of this method to guide experimental investigations of new ultra-refractory compositions will be the subject of further research.

A criterion for determining the eutectic point concentration in ultra-refractory alloys is proposed, which can narrow the search for eutectic in the expensive and time-consuming experimental investigations of new ultra-refractory compositions.

Formally, the neural network is not a mandatory element in the proposed algorithm. But first principle quantum calculations on samples of the desired size require an infeasibly large computational resources. The neural network, trained on the results of ''small`` first-principle calculations, and the subsequent molecular dynamics simulations solve this problem.

Note also that applying the algorithm to high entropy alloys ~\cite{Araujo24_JCP,Queiro25_JCP,Shang25_JCP,Zheng25_JCP,Wang23_nCM,Zhang24_JAP}
does not require a significant change in approach.

AI-approach based on the study of the properties of the corresponding melt on the pseudo-binary section of multi-component compositions makes it possible to determine the eutectic points with reasonable accuracy without the needs to know the crystal structure details of the pre- and post-eutectic solid alloys. This method can be used for design of advanced multicomponent ultra-refractory materials.

\section{Acknowledgements}

The computations were carried out on MVS-10P at Joint Supercomputer Center of the Russian Academy of Sciences (JSCC RAS). This work has been carried out using also computing resources of the Federal Collective Usage Center Complex for Simulation and Data Processing for Mega-Science Facilities at NRC ``Kurchatov Institute'', http://ckp.nrcki.ru/.
E.A.Levashov would like to acknowledge the financial support from the Russian Science Foundation (No.24-13-00085)



\end{document}